IAC-24,D5,IP,12,x89662

# A Machine Learning-Ready Data Processing Tool for Near Real-Time Forecasting

Maher A Dayeh[a,b] *, Michael J Starkey[a], Subhamoy Chatterjee[c], Heather Elliott[a,b], Samuel Hart[b,a], and Kimberly Moreland [e,d,b,a]

[a] *Southwest Research Institute, San Antonio, TX, USA; maldayeh@swri.edu*
[b] *The University of Texas at San Antonio, San Antonio, TX, USA*
[c] *Southwest Research Institute, Boulder, CO, USA*
[d] *Space Weather Prediction Center, NOAA, Boulder, CO, USA*
[e] *Cooperative Institute for Research in Environmental Sciences, University of Boulder, CO, USA*
\* Corresponding Author

**Abstract**

Space weather forecasting is critical for mitigating radiation risks in space exploration and protecting Earth-based technologies from geomagnetic disturbances. This paper presents the development of a Machine Learning (ML)-ready data processing tool for Near Real-Time (NRT) space weather forecasting. By merging data from diverse NRT sources such as solar imagery, magnetic field measurements, and energetic particle fluxes, the tool addresses key gaps in current space weather prediction capabilities. The tool processes and structures the data for machine learning models, focusing on time-series forecasting and event detection for extreme solar events. It provides users with a framework to download, process, and label data for ML applications, streamlining the workflow for improved NRT space weather forecasting and scientific research.

**Acronyms/Abbreviations**
- Solar and Heliospheric Observatory (SOHO)
- Large Angle and Spectrometric Coronagraph (LASCO)
  - Coronagraph 2 (C2)
  - Coronagraph 3 (C3)
- Extreme Ultraviolet Imaging Telescope (EIT)
- Michelson Doppler Imager (MDI)
- Solar Terrestrial Relations Observatory (STEREO)
- Sun Earth Connection Coronal and Heliospheric Investigation (SECCHI)
  - Coronagraph 2 (COR2)
  - Heliospheric Imager 1 (HI1)
  - Heliospheric Imager 2 (HI2)
- STEREO/WAVES (SWAVES)
- Deep Space Climate Observatory (DSCOVR)
- Faraday Cup (FC)
- Magnetometer (MAG)
- Solar Dynamics Observatory (SDO)
- Extreme Ultraviolet Variability Experiment (EVE)
- Atmospheric Imaging Assembly (AIA)
- Helioseismic and Magnetic Imager (HMI)
- Geostationary Operational Environmental Satellite (GOES)
- X-ray Sensor (XRS)
- Solar Ultraviolet Imager (SUVI)
- Energetic Proton Detector (Protons)
- Energetic Electron Detector (Electrons)
- Advanced Composition Explorer (ACE)
- Electron, Proton, and Alpha Monitor (EPAM)
- Solar Wind Electron, Proton, and Alpha Monitor (SWEPAM)
- Magnetometer (MAG)
- Space weather HMI Active Region Patch (SHARP) parameters
- National Solar Observatory / Global Oscillation Network Group (NSO/GONG)
- Planetary K-index (Kp Index)
- Planetary A-index (Ap Index)

## 1. Introduction

Space weather forecasting plays a vital role in space exploration and the protection of critical systems both in space and on Earth. The dynamic interactions between the Sun and Earth's magnetic field can cause adverse effects such as increased radiation levels in space and geomagnetic storms, posing significant risks to satellites, astronauts, and terrestrial infrastructure [1], [2], [3]. In this context, the prediction of space weather events, such as solar flares and coronal mass ejections (CMEs), requires robust data processing methods that can handle large volumes of NRT data from a wide range of sources. Recent advances in Machine Learning (ML) techniques provide opportunities to enhance the prediction of space weather events [4], [1], but a major challenge is the preparation of multi-instrument datasets in a format suitable for ML models. This paper presents the development of a Python-based data processing tool that integrates NRT data streams from multiple spacecraft and instruments into a unified, ML-ready format. The tool will significantly advance space weather forecasting by offering streamlined access to a diverse array of near-real-time datasets from a variety of active space- and ground-based resources [5], thus facilitating broader, comprehensive and timely space weather analyses and aiding in the development of strategies to mitigate potential risks associated with space weather events [6], [7].

**2. Tool Overview**

The tool is designed to prepare and process data from multiple NRT sources. It consists of three key modules:

- Data Downloader: This module retrieves NRT data from a variety of satellite missions such as SOHO, ACE, GOES, DSCOVR, and STEREO, among many others. Users can specify the data sources, time ranges, and instrument parameters to download historical data or stream NRT data directly. Table 1. Lists the main data sources and types.

Table 1. List of data products available in near real time

| # | Instrument/Source | NRT Data Type |
|---|---|---|
| 1 | SOHO/LASCO/C2 | Imagery 2.2-6 Rs |
| 2 | SOHO/LASCO/C3 | Imagery 3.7-30 Rs |
| 3 | SOHO/EIT | Imagery |
| 3 | SOHO/MDI ** | Imagery |
| 5 | STEREO/ SECCHI /COR2 | Imagery 2.5-15 Rs |
| 6 | STEREO/SECCHI/HI1 | Imagery 15-84 Rs |
| 7 | STEREO/SECCHI/HI2 | Imagery 66-318 Rs |
| 8 | STEREO/SWAVES | Type II Radio Burst |
| 9 | DSCOVR/FC | Solar Wind V,N,T |
| 10 | DSCOVR/MAG | IMF |
| 11 | SDO/EVE | EUV Irradiance |
| 12 | SDO/AIA | EUV, 17.1 nm |
| 13 | SDO/HMI | Magnetograms |
| 14 | GOES/XRS | Solar X-Ray Flux |
| 15 | GOES/SUVI | Solar EUV Flux |
| 16 | GOES/Protons | Energetic Particles |
| 17 | GOES/Electrons | Electrons |
| 18 | ACE/EPAM | Ions > 50keV |
| 19 | ACE/SWEPAM | Solar Wind V,N,T |
| 20 | ACE/MAG | IMF |
| 21 | SHARP parameters | Magnetograms |
| 22 | NSO/GONG | Magnetograms |
| 23 | Kp, Ap indices | Geomagnetic |

- Data Processor: Once downloaded, the data undergoes cleaning and processing to ensure consistency. The processor performs tasks such as time-stamp normalization, detection and correction of missing or degraded data, and identification of data outliers using statistical methods. Processed data is stored or can be exported in common formats (CSV for time-series data, FITS for images) for direct use in ML models.

- Data Splitter: Available in Historical Mode, the Data Splitter divides the processed dataset into training, validation, hold-out, and test sets, ensuring that ML models are trained on robust data partitions. Users can customize the split based on random draws, time intervals, or data clusters.

**3. Data Processing and Preparation**

The Data Processor module is central to the tool's ability to generate ML-ready datasets. It handles data in two main stages:

- Time Stamp Normalization: Various NRT sources provide data with different time formats. The tool standardizes all timestamps into a single format (e.g., Julian Date), enabling seamless data integration across different instruments.

- Missing Data and Degradation Correction: Space-based instruments often have gaps in data due to operational anomalies or other factors. The tool identifies and flags missing data points with consistent fill values. Degradation correction algorithms for specific instruments (e.g., SDO/AIA's EUV channel degradation) are integrated, leveraging existing libraries such as SunPy and AIApy for normalization.

- Data Outlier Detection: Two independent outlier detection algorithms are implemented:

  - The **InterQuartile Range (IQR) method** identifies outliers based on statistical dispersion within a defined time window, while iterating over the dataset to ensure accurate removal of outliers.
  - The **Z-score method** computes the z-score relative to the mean and standard deviation within a time window to flag data points significantly deviating from the expected range.

These techniques improve data quality for ML training by removing erroneous values that could skew model performance.

**4. Modes of Operation**

- Historical Mode: Users can specify a time range to download and process historical data, which is then divided into training, validation, and test sets using the Data Splitter. This mode is ideal for developing and training ML models that require extensive, curated datasets for event forecasting.
- NRT Streaming Mode: The tool continuously downloads and processes the latest NRT data, storing the processed dataset in a buffer of user-defined size. Users can set the tool to run at regular intervals (e.g., hourly updates), ensuring that the most recent space weather data is always available for real-time forecasting.

## 5. Implementation and Validation

The tool leverages the Snakemake workflow management system to ensure that the data processing pipeline is reproducible and scalable [8]. Snakemake allows for flexible configuration of user inputs, ensuring that the pipeline can handle varying datasets and operational modes. Validation flags are used to confirm the continuity and accuracy of data processing tasks. Figure 1 illustrates the pipeline of the tool.

*Figure 1. Pipeline for the tool concept of operation. The pipeline enables the user to download data that gets processed and ML-ready to feed into models of interest. The tool's NRT historic and NRT stream data modes enable model training and operating on NRT datasets, utilizing a large variety of in-situ and remote observational resources. VSO, DRMS, and Fido are Python library packages.*

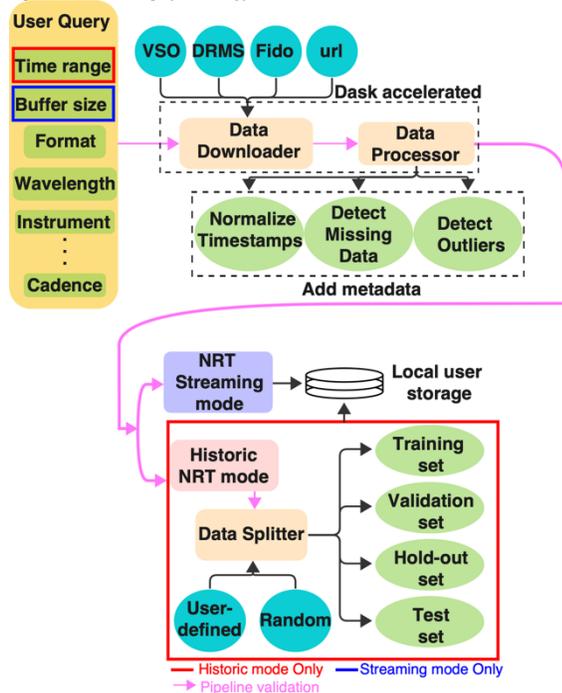

## 6. Summary

This paper introduces a comprehensive Python-based tool for the preparation of ML-ready datasets from NRT space weather data. By integrating data from multiple in-situ and remote sensing instruments, the tool fills an important gap in current space weather prediction capabilities. It streamlines the workflow for developing, training, and testing ML models aimed at predicting extreme solar events. Future work will involve expanding the tool's functionality to include additional data sources and real-time integration with operational forecasting models, and packaging it into a Python-ready library.

## Acknowledgements

Authors acknowledge partial support from NASA grants LWS (80NSSC19K0079, 80NSSC21K1316, 80NSSC20K1815, 80NSSC21K1307, and 80NSSC21K1324) and O2R (80NSSC21K0027)